\documentclass[aps,twocolumn,pra,superscriptaddress]{revtex4}
\usepackage{epsfig,graphicx,times}
\usepackage{amstext}
\usepackage{amsmath}
\usepackage{lipsum}
\usepackage{amssymb}
\usepackage{float}
\usepackage{graphicx}
\usepackage{latexsym}
\usepackage{bm}
\usepackage{epstopdf}
\usepackage{appendix}
\usepackage[colorlinks,citecolor=blue, linkcolor=blue,hyperindex,CJKbookmarks]{hyperref}

\begin{document}

\title{Aging of coupled qubits}
\author{Huining Zhang}
\affiliation{Center for Quantum Sciences and School of Physics, Northeast Normal University, Changchun 130024, China}
\author{Dianzhen Cui}
\affiliation{Center for Quantum Sciences and School of Physics, Northeast Normal University, Changchun 130024, China}
\author{W. Wang}
\affiliation{ School of Physics, Northeast Normal University, Changchun 130024, China}
\author{X. X. Yi\footnote{yixx@nenu.edu.cn}}
\affiliation{Center for Quantum Sciences and School of Physics, Northeast Normal University, Changchun 130024, China}
\affiliation{Center for Advanced Optoelectronic Functional Materials Research, and Key Laboratory for UV Light-Emitting Materials and Technology of Ministry of Education, Northeast Normal University, Changchun 130024, China}

\date{\today}

\begin{abstract}
The aging transition refers to the shift from an oscillatory state to a globally ceased state due to some forms of deterioration in classical physics.
Similar behavior has also been observed in quantum oscillators.
Although it has received extensive attention in coupled oscillator systems, it has not yet been studied in coupled qubits.
In this manuscript, we explore the aging transition in a network of coupled qubits.
Our model describes {numerous} qubits driven by a laser, with both dissipative and coherent qubit-qubit couplings.
The ratio of inactive qubits to total qubits and the population in the excited state of the qubits are employed to characterize the aging transition.
We find a transition where the population in the excited states suddenly drops when the ratio exceeds a threshold.
This behavior is intriguing and contrasts with coupled oscillators, where no sudden drop is observed.
Additionally, we demonstrate how the couplings and driving laser influence the threshold.
The underlying physics of the sudden drop is elucidated.
The region where the aging transition occurs is determined based on stability analysis theory.
\end{abstract}
\maketitle

\section{INTRODUCTION}

Aging is the time-related deterioration of the physiological functions necessary for survival and fertility,  occurring as living organisms degrade with age. In physics, aging is considered a collective dynamical process in a network of coupled oscillators \cite{Daido2004,Bandy2023,Daido2007,Daido2008,Sun2019,Sun2017,Sath2019,Sath2022,Ponrasu2020,Biswas2022,Sahoo2023,Singh2020,Rahman2017,Rakshit2020,Zou2021}. When some oscillators become non-self-oscillatory,  the network is said to be aging \cite{Daido2004}. The aging of coupled oscillator network was first considered in Ref. \cite{Daido2004}, in which some oscillators are in inactive states while the others are in active states. Here, inactive and active states  refer to non-self-oscillatory and self-oscillatory states, respectively.  It was illustrated  that as the ratio of inactive oscillators to the  total number of oscillators increases,  an aging transition occurs, causing the oscillations in the network's dynamics to disappear.

Recently, it has been shown that the aging transition can also occur in quantum coupled oscillators, but its features are distinctly different from those of its classical counterpart \cite{Bandy2023}.
For the quantum oscillators network, the active and inactive oscillators are defined based on the dissipative properties of the oscillator.
If the oscillators are expressed in terms of boson representation, the aging transition cannot be marked by the complete collapse of the system but by a rapid decrease in the mean boson number.
There is a ``knee" in the dependence of the mean boson number on the ratio of inactive to total oscillators. Although significant progress has been made on the aging transition \cite{Daido2004,Bandy2023,Daido2007,Daido2008,Sun2019,Sun2017,Sath2019,Sath2022,Ponrasu2020,Biswas2022,Sahoo2023,Singh2020,Rahman2017,Rakshit2020,Zou2021}, a discussion on the aging transition problem for coupled qubits is rarely addressed.

In this manuscript, we focus on the aging transition in a network of coupled qubits and present the differences between aging transitions in networks with qubits and those with oscillators. Coupled qubit systems play an important role in quantum information and quantum simulations \cite{Niel2010,Imam1999,Monroe1995,Chio2004}. Due to unavoidable couplings of the system to its surroundings, all systems experience some form of aging or deterioration in the real world, which destroys system activity in various ways. Examples include lossy cavities \cite{Yuge2014}, trapped ions with ion-ion collisions \cite{peder2002}, optomechanical systems with mechanical dissipation \cite{Aspel2014,Fitzg2021}, and dielectric superconducting junctions with decoherence \cite{Blais2021}. Studying the aging of coupled qubits may help determine how many inactive qubits need to be revitalized for a quantum device to function properly.

Here, we consider a network composed of $N$ coupled active and inactive qubits driven by a laser with global coherent and dissipative couplings.
The active and inactive qubits are defined based on the characteristics of the dissipators in the quantum master equation. By adjusting the ratio of inactive qubits $p$, we demonstrate that the mean excited-state population undergoes a sudden drop when this ratio exceeds a threshold, which precisely characterizes the aging transition.
Furthermore, we illustrate that the coherent coupling, dissipative coupling, and the driving laser each play distinct roles in the aging transition.
We  identify a parameter region where the mean excited-state population exhibits two stable solutions as a function of $p$, leading to the aging transition.
Analytically determining the boundaries between monostable and bistable regimes allows us to pinpoint the interval within which the aging transition occurs, and the extent of this interval can be adjusted by tuning the system parameters.

This work is organized as follows. In Sec.\ref{sec1}, we introduce our model and employ two approaches to calculate the population in the excited states of the system.
In Sec.\ref{sec2}, we present the main results of this work. We reveal the physics behind the observation of a transition where the population in the excited states undergoes a sudden drop, and we discuss the implications of these results.
A brief discussion on the experimental realization of our model concludes this section.
Finally, in Sec.~\ref{sec3}, we summarize our findings. Appendix \ref{appendixA} provides details for the derivation of the equations in the collective motion approach.
Appendix \ref{appendixB} compares the results from the collective motion approach, mean-field theory, and exact numerical simulations.
Appendix \ref{appendixC} analyzes the difference between the results given by equations of collective motion and mean-field theory.
Appendix \ref{appendixD} discusses the stability analysis for each fixed point, and Appendix \ref{appendixE} presents the derivation of the equations for collective motion in the presence of quantum correlations.

\section{MODEL AND METHOD}\label{sec1}

In this section, we first present our model and introduce a parameter to characterize the aging transition.
We employ two approaches to address the problem.
The first approach involves treating the system collectively, where we derive a set of equations for the collective operators of the coupled qubits.
The second approach utilizes mean-field theory.
In contrast to the collective approach, this method analyzes the aging transition by examining individual qubits within the system.

For $N$ dissipative qubits with global coupling and laser driving, the Hamiltonian reads (with $\hbar=1$)
\begin{equation}
H = \sum_{j=1}^N \frac{\Delta}{2} \sigma_{z}^{j} + \sum_{j=1}^N \frac{\Omega}{2} (\sigma_{+}^{j} + \sigma_{-}^{j}) +{\sum_{j,k}}^{'}g\sigma_{z}^{j}\sigma_{z}^{k},
\label{Hamiltonian}
\end{equation}
where the terms denote: the free Hamiltonian of the system with detuning $\Delta = \omega_0 - \omega_l$, where $\omega_0$ is the transition frequency of each qubit and $\omega_l$ is the laser frequency; laser-qubit coupling and coherent qubit-qubit coupling with strengths $\Omega$ and $g$, respectively. The sum ${\sum_{j,k}}^{'}$ indicates summation over distinct qubit indices. Here, $\sigma_{z}^{j} = |e_j\rangle\langle e_j| - |g_j\rangle\langle g_j|$, $\sigma_{+}^{j} = |e_j\rangle\langle g_j|$, and $\sigma_{-}^{j} = |g_j\rangle\langle e_j|$ are Pauli operators, with $|e_j\rangle$ and $|g_j\rangle$ denoting the excited and ground states of the $j$-th qubit. We adopt the following master equation to describe the system's dynamics \cite{Bandy2023,Lee2014,Ishi2017},
\begin{eqnarray}
\dot{\rho} &=& -i[H,\rho] + {\sum_{j,k}}^{'} \frac{V}{N} \mathcal{D}[\sigma_{-}^{j} - \sigma_{-}^{k}](\rho) \nonumber \\
&& + \sum_{j=1}^{N_a} \kappa_{1j} \mathcal{D}[\sigma_{+}^{j}](\rho) + \sum_{j=N_a+1}^{N} \kappa_{2j} \mathcal{D}[\sigma_{-}^{j}](\rho),
\label{masterequation}
\end{eqnarray}
where $\rho$ denotes the system's density matrix, and $\mathcal{D}[\hat{L}](\rho) = 2\hat{L}\rho\hat{L}^{\dagger} - \{\hat{L}^{\dagger}\hat{L},\rho\}$.
The second term describes joint dissipations inducing qubit-qubit couplings with strength $V$, referred to as dissipative coupling.

The last two terms \(\mathcal{D}[\sigma_{+}^{j}]\) and \(\mathcal{D}[\sigma_{-}^{j}]\) in the quantum master equation distinguish between active and inactive qubits.
Active qubits undergo incoherent transitions to the state \(|e_j\rangle\) from \(|g_j\rangle\) with rates \(\kappa_{1j}\), while inactive qubits transition from \(|e_j\rangle\) to \(|g_j\rangle\) with rates \(\kappa_{2j}\).
These definitions align with those used for quantum oscillators \cite{Bandy2023}.
With this formulation, \(N\) coupled qubits are partitioned into two groups: \(N_a\) active qubits and \(N_i = N - N_a\) inactive qubits, resulting in the ratio of inactive qubits \(p = \frac{N_i}{N} = \frac{N - N_a}{N}\).
In what follows, active qubits are denoted by \(j \in \{1, \ldots, N(1-p)\}\), while inactive qubits are denoted by \(j \in \{N(1-p)+1, \ldots, N\}\).

In order to characterize the aging of qubits, we define
\begin{equation}
\overline{n}(p) = \langle Q\rangle = \frac{1}{N} \sum_j \langle \sigma_{+}^{j} \sigma_{-}^{j} \rangle,
\label{order}
\end{equation}
where $\langle Q\rangle = \text{Tr}(\rho Q)$ with $Q = \frac{1}{N} \sum_{j} \sigma_{+}^{j} \sigma_{-}^{j}$. For large $N$, calculating $\overline{n}(p)$ directly is challenging.
To address this challenge, we assume equal dissipative rates for all qubits, i.e., $\kappa_{1j} = \kappa_{2j} = \kappa$.
Under this assumption, the evolution equation for $\langle Q\rangle$ becomes
\begin{eqnarray}
\dot{\langle Q\rangle} &=& \frac{-i\Omega}{2N} \sum_{j=1}^{N} (\langle \sigma_{+}^{j} \rangle - \langle \sigma_{-}^{j} \rangle) + 2\kappa(1-p) \nonumber \\
&& + \frac{2V}{N^2} {\sum_{j,k}}^{'} (\langle \sigma_{+}^{j} \sigma_{-}^{k} \rangle + \langle \sigma_{+}^{k} \sigma_{-}^{j} \rangle) \nonumber \\
&& - (2\kappa + \frac{4V(N-1)}{N}) \langle Q\rangle.
\label{collective operator1}
\end{eqnarray}
To obtain a closed set of equations, we derive the dynamical equation for the operator $A = \frac{1}{N} \sum_{j} \sigma_{+}^{j}$,
\begin{eqnarray}
\dot{\langle A\rangle} &=& i\Delta \langle A\rangle - 4ig(N-1) \langle A\rangle - \kappa \langle A\rangle \nonumber \\
&& - \frac{i\Omega}{2} (2\langle Q\rangle - 1) + \frac{4ig}{N} {\sum_{j,k}}^{'} 2 \langle \sigma_{+}^{j} \sigma_{+}^{k} \sigma_{-}^{k} \rangle \nonumber \\
&& - \frac{4V}{N^2} {\sum_{j,k}}^{'} \langle \sigma_{+}^{j} \sigma_{-}^{j} \sigma_{+}^{k} \rangle.
\label{collective operator2}
\end{eqnarray}
For a detailed derivation of Eq.(\ref{collective operator1}) and Eq.(\ref{collective operator2}), please refer to Appendix \ref{appendixA}.

Next, we assume the couplings among the qubits are weak and the number of qubits $N$ is large. These considerations ensure the following approximations
\begin{eqnarray}
{\sum_{j,k}}^{'}\langle\sigma_{+}^{j}\sigma_{-}^{j} \sigma_{+}^{k}\rangle&\rightarrow&{\sum_{j,k}}\langle\sigma_{+}^{j}\sigma_{-}^{j}\rangle\langle\sigma_{+}^{k}\rangle,\nonumber\\
{\sum_{j,k}}^{'}\langle\sigma_{+}^{j}\sigma_{-}^{k}\rangle&\rightarrow&{\sum_{j,k}}\langle\sigma_{+}^{j}\rangle\langle\sigma_{-}^{k}\rangle,
\label{approximation}
\end{eqnarray}
where the right-hand side terms differ from the left-hand side of  Eq.~(\ref{approximation}) due to the inclusion of $j=k$, acknowledging negligible correlations between different qubit pairs.
With these approximations, we derive a set of three nonlinear differential equations for $\langle Q \rangle$, $\langle A \rangle$, and ${\langle A \rangle}^\ast$ as
\begin{eqnarray}
\label{zhankaisuoxie1}
\dot{\langle Q\rangle} &=& \Omega {\rm Im}\langle A\rangle - (2\kappa + \frac{4V(N-1)}{N})\langle Q\rangle + 2\kappa(1-p)\nonumber\\
&& + 4V{|\langle A\rangle|}^2, \nonumber\\
\dot{\langle A\rangle} &=& i[\Delta - 4g(N-1) + 8gN\langle Q\rangle]\langle A\rangle + \frac{i\Omega}{2}(1 - 2\langle Q\rangle)\nonumber\\
&&  - (\kappa + 4V\langle Q\rangle)\langle A\rangle, \nonumber\\
\dot{{\langle A\rangle}^\ast} &=& (\dot{{\langle A\rangle}})^\ast.
\end{eqnarray}

\begin{figure}[t]
	\centering
	\includegraphics[width=0.45\textwidth]{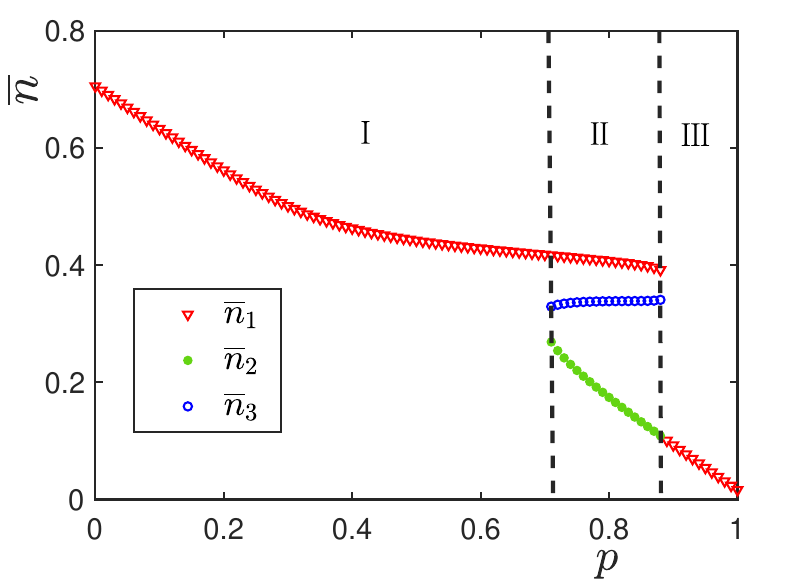}
	\caption {Fixed points of $\overline{n}$ as a function of the ratio of the inactive qubits $p$. The figure is divided into three regions. In regions \uppercase\expandafter{\romannumeral1} and \uppercase\expandafter{\romannumeral3}, $\overline{n}$ has only one fixed point $\overline{n}_1$ since $(\frac{m}{2})^2+(\frac{n}{3})^3>0$. In region \uppercase\expandafter{\romannumeral2}, there are three fixed points $\overline{n}_1, \overline{n}_2$, and $\overline{n}_3$ since $(\frac{m}{2})^2+(\frac{n}{3})^3<0$. The  parameters chosen are $V$=0.2$\kappa$, $N$=100, $\Delta$=3$\kappa$, $g$=0.04$\kappa$, and $\Omega$=3.2$\kappa$.}
	\label{aqbpic1}
\end{figure}

Because Eqs. (\ref{zhankaisuoxie1}) are nonlinear, the solution may not be uniquely determined.
We analyze the steady-state solutions (fixed points) by solving $\dot{\langle Q\rangle} = \dot{\langle A\rangle} = \dot{{\langle A\rangle}^\ast} = 0$, which leads to
\begin{eqnarray}
\begin{aligned}
\langle A\rangle = \frac{i\Omega(2\langle Q\rangle - 1)}{2i\Delta - 8ig(N-1) - 2\kappa + (16igN - 8V)\langle Q\rangle}.
\label{analytic}
\end{aligned}
\end{eqnarray}
Through simple calculations, we can obtain
\begin{eqnarray}\label{Qeq}
\begin{aligned}
a\langle Q\rangle^3 + b\langle Q\rangle^2 + c\langle Q\rangle + d = 0,
\end{aligned}
\end{eqnarray}
with
\begin{eqnarray*}
a &=& [2\kappa + \frac{4V(N-1)}{N}](64g^2N^2 + 16V^2),\\
b &=& [2\kappa + \frac{4V(N-1)}{N}][16gN(\Delta - 4g(N-1)) + 8\kappa V]\\
&& - 2\kappa(64g^2N^2 + 16V^2)(1-p),\\
c &=& [2\kappa + \frac{4V(N-1)}{N}][(\Delta - 4g(N-1))^2 + \kappa^2]\\
&& - 2\kappa[16gN(\Delta - 4g(N-1)) + 8\kappa V](1-p) \\
&&+ \Omega^2(\kappa + 2V),\\
d &=& -\frac{\Omega^2}{2}(\kappa + 2V) - 2\kappa[(\Delta - 4g(N-1))^2 + \kappa^2](1-p).
\end{eqnarray*}
Since the mean population in the excited states $\overline{n}(p) = \frac{1}{N} \sum_j \langle\sigma_{+}^{j}\sigma_{-}^{j}\rangle = \frac{1}{N} \sum_j {|{\langle e_j|\psi\rangle}|}^2$ is not negative,  the fixed point of $\overline{n}(p)$ must be real, which leads to discard all complex solutions of  Eq.~(\ref{Qeq}).
The solution of Eq.~(\ref{Qeq}) reads
\begin{eqnarray}
\overline{n}_1(p) &=& \alpha + \beta - \frac{b}{3a}, \nonumber\\
\overline{n}_2(p) &=& \omega \alpha + \omega^2 \beta - \frac{b}{3a}, \nonumber\\
\overline{n}_3(p) &=& \omega^2 \alpha + \omega \beta - \frac{b}{3a},
\label{roots}
\end{eqnarray}
where $\alpha=\big[-\frac{m}{2}+ \sqrt{(\frac{m}{2})^2+(\frac{n}{3})^3}\big]^\frac{1}{3}$, $\beta=\big[-\frac{m}{2}- \sqrt{(\frac{m}{2})^2+(\frac{n}{3})^3}\big]^\frac{1}{3},$ $m=\frac{d}{a}-\frac{bc}{3a^2}+\frac{2}{27}(\frac{b}{a})^3$, $n=\frac{c}{a}-\frac{1}{3}(\frac{b}{a})^2$ and $\omega=\frac{-1+\sqrt{3}i}{2}$.
For $\left(\frac{m}{2}\right)^2 + \left(\frac{n}{3}\right)^3 > 0$, there is only one fixed point $\overline{n}_1(p)$ for $\overline{n}$, which attributes to that the other two solutions of Eq.~(\ref{analytic}) are complex.
For $\left(\frac{m}{2}\right)^2 + \left(\frac{n}{3}\right)^3 < 0$, $\overline{n}(p)$ has three fixed points, denoted by $\overline{n}_1(p)$, $\overline{n}_2(p)$, and $\overline{n}_3(p)$, and they are all real. The dependence of the fixed points on $p$ is shown in FIG.~\ref{aqbpic1}, in which we include the terms where $j=k$ and ignore the correlation between qubits.

To verify the effectiveness of the collective operator method, we compare our results with those obtained using the mean-field approach.
In mean-field theory \cite{Diehl2010,Lee2011,Tomadin2011}, the density matrix for the whole system approximates to a product of the density matrices of all subsystems, $\rho \approx \bigotimes_j \rho_j$. This approximation is effective for weak qubit-qubit couplings, where correlations among the qubits can be neglected. Applying this approximation to Eq. (\ref{masterequation}), we obtain a reduced master equation for the $j$th density matrix $\rho_j = \text{Tr}_{\neq j}\rho$, given by
\begin{eqnarray*}
\begin{aligned}
\dot{\rho_j} &= -i[\frac{\Delta}{2}\sigma_z^j + \frac{\Omega}{2}(\sigma_+^j + \sigma_-^j) + 2g\sigma_z^j {\sum_{jk}}^{'} (2w_k - 1), \rho_j] \\
&  + \kappa \mathcal{D}[\hat{L}_j](\rho_j)  + \frac{2V}{N}({\sum_{jk}}^{'}q_k^* [\sigma_+^j, \rho_j] + {\sum_{jk}}^{'} q_k [\rho_j, \sigma_-^j]) \\
&  + \frac{2V(N-1)}{N} \mathcal{D}[\sigma_-^j](\rho_j),
\end{aligned}
\end{eqnarray*}
where we define the excited-state population of the $j$th qubit as $w_j \equiv \langle e_j | \rho_j | e_j \rangle$, and the off-diagonal element of the density matrix as $q_j \equiv \langle g_j | \rho_j | e_j \rangle$. $\hat{L}_j = \sigma_+^j$ represents the Lindblad operators for active qubits, while $\hat{L}_j = \sigma_-^j$ represents those for inactive qubits. The evolution of $w_j$ and $q_j$ is given by
\begin{equation}
\begin{aligned}
\dot{w_j} &=
\begin{cases}
\Omega \mathrm{Im} q_j - (2\kappa + \frac{4V(N-1)}{N}) w_j + 2\kappa \\+ \frac{2V}{N} ( q_j \sum_k' q_k^* + q_j^* \sum_k' q_k ), & \text{active qubits} \\
\\
\Omega \mathrm{Im} q_j - (2\kappa + \frac{4V(N-1)}{N}) w_j \\+ \frac{2V}{N} ( q_j \sum_k' q_k^* + q_j^* \sum_k' q_k), & \text{inactive qubits}
\end{cases}
\\
\dot{q_j} &= i[\Delta + 4g {\sum_k}^{'} (2w_k - 1)] q_j + i \frac{\Omega}{2} (1 - 2w_j) - \kappa q_j \\
&- \frac{4V}{N} {\sum_k}^{'} w_j q_k - \frac{2V(N-1)}{N} q_j + \frac{2V}{N} {\sum_k}^{'} q_k.
\end{aligned}
\label{wj}
\end{equation}
\begin{figure*}[t]
\centering
\includegraphics[width=5.2cm]{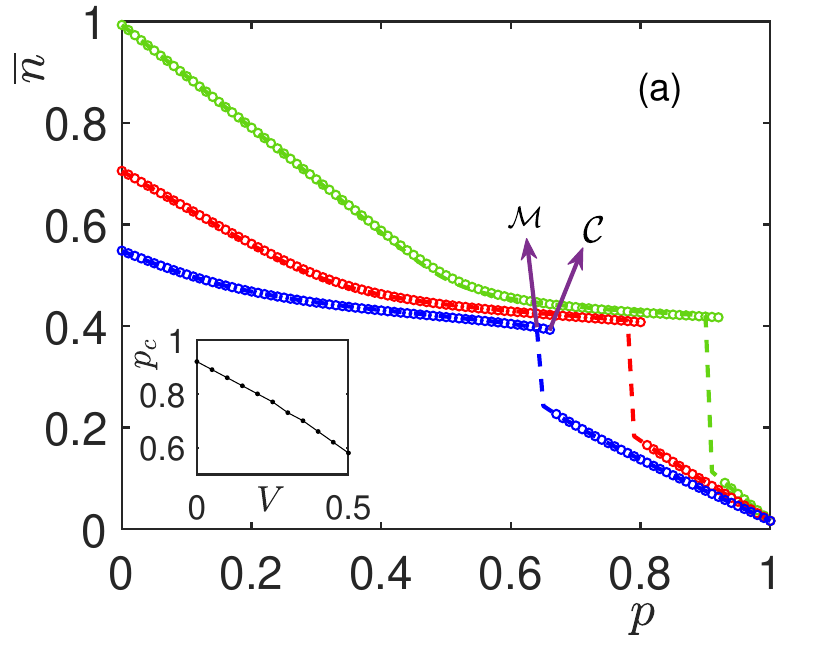}
\includegraphics[width=5.6cm]{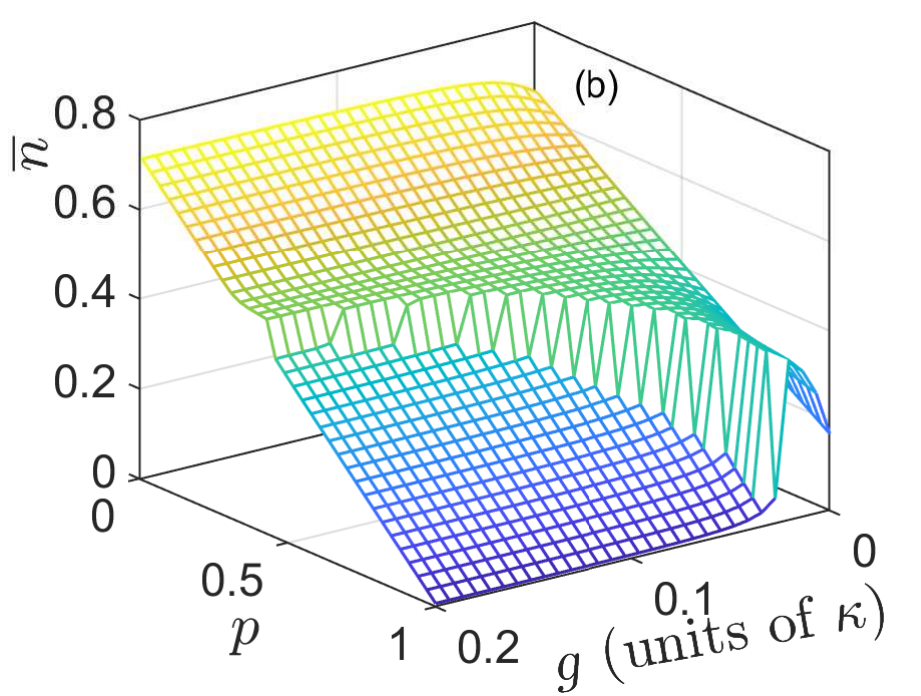}
\includegraphics[width=5.6cm]{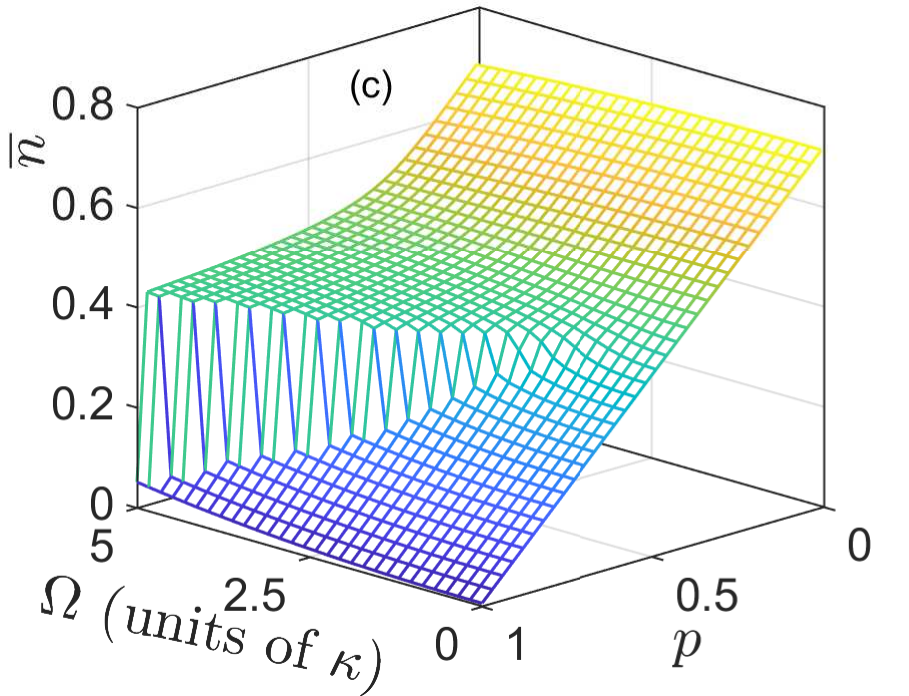}
\caption{(a) The mean excited-state population versus the ratio of inactive qubits for different dissipative coupling strengths $V$ in the two methods. There are 6 lines (except the inset) in figure (a)  in 3 different color. $V$ = 0, 0.2$\kappa$, 0.4$\kappa$ are for the green, red, and blue lines, respectively. The dashed lines are plotted by the mean-field theory and the circle lines are plotted by solving the equations of collective motion. $\mathcal{M}$ marks the critical point $p_c$ given by the mean-field theory, while $\mathcal{C}$  the critical point $p_c$ given by the equation of collective motion. (b) $\overline{n}$ as a function of $g$ and $p$ with $\Omega=3.2\kappa$. (c) $\overline{n}$ versus $\Omega$ and $p$ with $g=0.04\kappa$. The initial conditions for $\langle Q\rangle$, $\langle A\rangle$, and ${{{\langle A\rangle}}^\ast}$ are denoted by $\langle Q\rangle_0$=0.5 and ${\langle A\rangle}_0={{{\langle A\rangle}_0}^\ast}=0.5$. Other parameters chosen are the same as in FIG.~\ref{aqbpic1}.}
\label{aqbpic234}
\end{figure*}
It has been demonstrated that the results derived from both the collective motion equations and mean-field theory align well with exact solutions, as detailed in Appendix \ref{appendixB}.
Mean-field theory yields a set of \(3N\) closed nonlinear differential equations.
However, solving this extensive set of equations becomes time-consuming for large numbers of qubits in comparison to the collective motion equations.

\section{RESULTS AND DISCUSSION}\label{sec2}

In the preceding section, we introduced two sets of equations based on collective motion and mean-field theory, respectively.
In this section, we numerically solve these equations, analyze the characteristics of the aging transition, and compare the results obtained from both methods.

In order to verify the existence of the aging transition, FIG.~\ref{aqbpic234}(a) illustrates $\overline{n}$ as a function of the ratio $p$ for different values of $V$.
Notable observation is the abrupt drop in $\overline{n}$, and there is a critical inactive qubits ratio \( p_c \) corresponding to this drop. This indicates that in a qubit network,  if the number of inactive elements exceeds a threshold, the mean excited-state population can sharply decline to a low value, contrasting with networks of coupled oscillators \cite{Daido2004,Bandy2023}.
We define the critical inactive qubits ratio as the aging transition point.
Another difference between qubits and oscillators lies in the condition for the aging transition. For oscillators, the dissipative coupling \( V \) must be non-zero and must exceed a threshold for the aging transition to occur \cite{Daido2004,Bandy2023}.
However, for qubits, the aging transition can occur even at \( V = 0 \).
As shown by the inset in FIG.~\ref{aqbpic234}(a), \( p_c \) decreases nearly linearly with increasing \( V \).
The third difference between qubit aging and oscillator aging is the behavior of the parameter on either side of the critical value.
For quantum oscillators, there are two characteristic zones.
Initially, the parameter decreases rapidly with the ratio \( p \). When \(p\) is larger than the critical value \( p_c \), the curve becomes almost a straight inclined line \cite{Bandy2023}. In contrast, for qubits, the slope of the curve following the aging transition is analogous to the slope observed when \(p\) is very small, which is illustrated by FIG.\ref{aqbpic234}(a).

As shown in FIG.~\ref{aqbpic234}(a), there is a slight difference in the critical value \( p_c \) predicted by the mean-field theory and the equation of collective motion.
This difference can be understood from the details provided in Appendix \ref{appendixC}.
Despite the slight deviation in the predicted critical value, the overall shapes of the curves are almost identical, indicating that the numerical results obtained from the two methods are not significantly different.
Given that mean-field theory involves substantial computational effort, whereas the collective motion equations offer simpler forms and facilitate analytical solutions, we employ the collective motion approach to further explore the influence of system parameters on aging transitions.

To  investigate the effect of the coherent coupling \( g \) and laser driving \( \Omega \) on the emergence of aging transitions in qubits, we plot \( \overline{n} \) as a function of \( g \) and \( p \) [FIG.~\ref{aqbpic234}(b)], as well as vs \( \Omega \) and \( p \) [FIG.~\ref{aqbpic234}(c)].
From FIG.~\ref{aqbpic234}(b), it is evident that beyond a certain threshold value of \( g \), an aging transition occurs where \( \overline{n} \) sharply decreases.
Interestingly, both dissipative coupling mediated by the environment and coherent coupling play analogous roles in facilitating the aging transition.
Stronger couplings lead to a smaller critical point \(p_c\), regardless of whether they are coherent or dissipative mechanisms.
To illustrate further, for a fixed \( p = 0.8 \), \( \overline{n} \) initially increases with \( g \).
Once \( g \)  surpasses a critical threshold, \( \overline{n} \) undergoes a sharp drop to a very low value.
As \( g \) continues to increase beyond this threshold, \( \overline{n} \) stabilizes towards a constant value.

In FIG.~\ref{aqbpic234}(c), the relationships between \( \overline{n} \), the driving strength \( \Omega \), and the ratio \( p \) are considered.
It becomes evident that the driving strength plays a crucial role in the qubit aging transition.
Specifically, when \( \Omega \) is below a certain threshold, the aging transition does not occur.
As \( \Omega \) increases, the critical point \( p_c \) shifts towards 1. This indicates that the driving force can influence the onset of aging transition. Furthermore, a stronger driving force amplifies the drop in \( \overline{n} \) at the critical point.
Choosing \( p = 0.8 \), the results illustrate that once \( \Omega \) exceeded a critical value, \( \overline{n} \) exhibits a significant increase followed by stabilization.
This ability to modulate \( \overline{n} \) through adjustments in coupling strength and laser driving provides a means to either enhance or suppress the mean excited-state population of qubit networks, assuming a fixed number of inactive qubits.
Overall, these findings emphasize the possibility of using couplings and external driving mechanisms to control and improve the performance of quantum systems affected by aging transitions.

The abrupt change in the parameter \( \overline{n} \) at the critical point \( p_c \) can be understood through an analysis of fixed points and their stability. In FIG.~\ref{aqbpic1}, we identify three fixed points labeled as \( \overline{n}_1 \), \( \overline{n}_2 \), and \( \overline{n}_3 \).
The stability of these fixed points is examined using linear stability analysis, detailed in Appendix \ref{appendixD}.
The analysis reveals that \( \overline{n}_1 \) is stable across the entire range of \( p \) from 0 to 1.
In contrast, \( \overline{n}_2 \) exhibits stability only within a specific region, referred to as region II in FIG.~\ref{aqbpic1}, while \( \overline{n}_3 \) is consistently unstable within region II.
The bistability observed in region II suggests that the system can settle into either \( \overline{n}_1 \) or \( \overline{n}_2 \) for a fixed \( p \), depending crucially on the initial conditions \( {\langle Q\rangle}_0 \) and \( {\langle A\rangle}_0 \).
Furthermore, owing to this bistability, even a slight shift in the ratio of inactive qubits \( p \) can significantly influence the fixed point that the system ultimately converges to.
This sensitivity highlights the crucial role of \( p \) in determining the system's final state and elucidates the origin of the sharp transition observed in \( \overline{n} \) at \( p_c \).

\begin{figure}[t]
	\centering
	\includegraphics[width=0.5\textwidth]{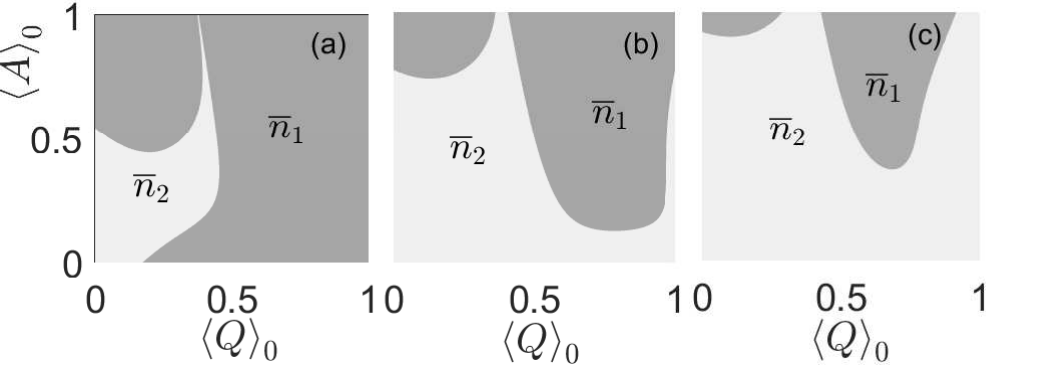}
	\caption{The stable regions  of the fixed points $\overline{n}_{1,2}$ versus ${\langle Q\rangle}_0$ and ${\langle A\rangle}_0$. $p$ is chosen in the bistable region (i.e., $p$ ranging from 0.71 to 0.88). Figure (a), (b), and (c) correspond to different $p$. (a) $p$ = 0.71,  (b) $p$ = 0.8, and (c) $p$=0.88. Other parameters chosen are the same as FIG.~\ref{aqbpic1}.}
	\label{aqbpic6}
\end{figure}

In FIG.~\ref{aqbpic6}, we illustrate the stable regions of the fixed points \( \overline{n}_{1,2} \) as a function of the initial conditions \( {\langle Q\rangle}_0 \) and \( {\langle A\rangle}_0 \), for several fixed values of \( p \).
The dark-gray regions indicate parameter sets where the system tends to evolve towards a larger \( \overline{n} \), stabilizing at \( \overline{n}_1 \).
Conversely, in the light-gray regions, the system stabilizes at a smaller \( \overline{n} \), corresponding to \( \overline{n}_2 \).
The boundary line separating fixed points \( \overline{n}_1 \) and \( \overline{n}_2 \) is determined by specific initial conditions.
The results highlight that the stable region of the fixed points \( \overline{n}_1 \) and \( \overline{n}_2 \) varies with \( p \).
Notably, the stable region of \( \overline{n}_1 \) diminishes as \( p \) increases, indicating a preference for the system to settle into the state with the lower mean excited-state population as the increase of the inactive qubits.
Importantly, the bistable region indicates that the aging transition can occur at any \( p \) within this region, provided the initial conditions are carefully chosen.
Here, \( p_{\text{cmin}} \) denotes the left boundary where the system first undergoes aging transition, while \( p_{\text{cmax}} \) represents the right boundary and the maximum \( p_c \) value for bistability.
This insight underscores the control one can exert over the onset of aging transition by manipulating initial conditions, thereby influencing the final state of the system.

\begin{figure}[t]
	\centering
	\includegraphics[width=0.45\textwidth]{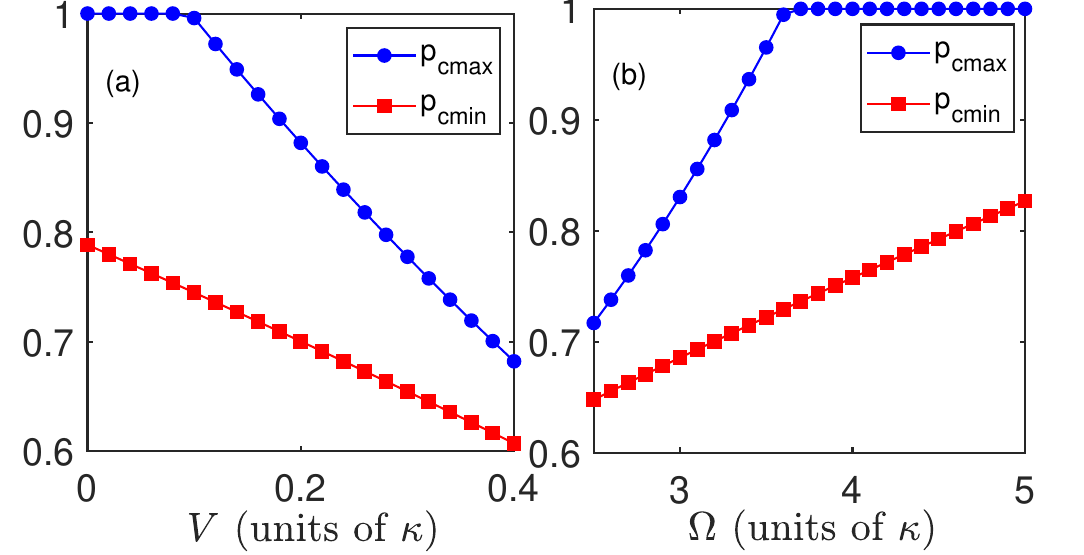}
	\caption {(a) The interval [$p_{\text{cmin}}$, $p_{\text{cmax}}$] for various dissipative coupling strength $V$ with $\Omega$=3.2$\kappa$. (b) The interval [$p_{\text{cmin}}$, $p_{\text{cmax}}$] for various laser driving strength $\Omega$ with $V$=0.2$\kappa$. Other parameters chosen are the same as FIG.~\ref{aqbpic1}.}
	\label{aqbpic7}
\end{figure}

As previously mentioned, \( p_{\text{cmin}} \) and \( p_{\text{cmax}} \) denote the boundaries of the bistable and monostable region, which determine where aging transitions occur. Therefore, it is valuable to investigate how \( p_{\text{cmin}} \) and \( p_{\text{cmax}} \) depend on \( V \) and the laser driving strength \( \Omega \).
FIG. \ref{aqbpic7}(a) illustrates that for smaller \( V \), \( p_{\text{cmax}} \) can reach 1, and the interval where aging transitions can occur expands with increasing \( V \); however, this interval subsequently narrows.
Regarding the driving strength \( \Omega \), the variation of \( p_{\text{cmin}} \) and \( p_{\text{cmax}} \) displayed in FIG. \ref{aqbpic7}(b) is symmetric with respect to FIG. \ref{aqbpic7}(a). The potential range for locating the aging transition point initially expands with increasing \( \Omega \); however, this range narrows when \( \Omega \) reaches a certain value. Unlike \( V \), \( p_{\text{cmax}} \) can reach 1 for larger \( \Omega \).
Moreover, the dynamics of the system are characterized by the upper and lower bounds of the critical point \( p_c \).
If  \( p \) is less than \( p_{\text{cmin}} \), the network cannot undergo an aging transition; if \( p \) is greater than \( p_{\text{cmax}} \), the aging transition has occurred, and the system is in a less active state.
\begin{figure}[b]
	\centering
	\includegraphics[width=0.43\textwidth]{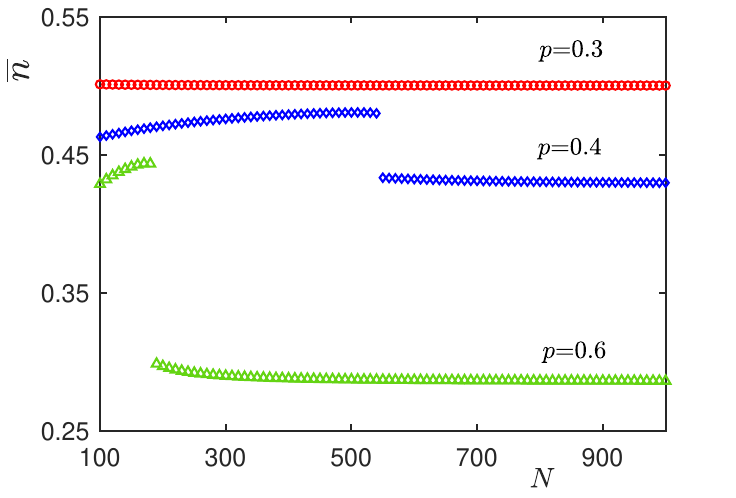}
	\caption{The parameter versus the total number of qubits $N$ for  several fixed $p$. $p$=0.3, 0.4, 0.6 from the top to down. ${\langle Q\rangle}_0$=0.5 and ${\langle A\rangle}_0$=${{\langle A\rangle}_0}^\ast$=0.5. Other parameters are chosen as  the same as in FIG.~\ref{aqbpic1}.}
	\label{aqbpic8}
\end{figure}

To investigate the behavior of the parameter \( \overline{n} \) for  a fixed ratio of inactive qubits and a variant total number of qubits, we solve the equations of collective motion and present the results in FIG.~\ref{aqbpic8}.
For a small fraction of inactive qubits (e.g., \( p = 0.3 \)), \( \overline{n} \) remains constant and independent of the total number \( N \) of qubits.
However, the situation changes as \( p \) increases (e.g., \( p = 0.4 \)). There exists a critical threshold in \( N \), beyond which the mean excited-state population \( \overline{n} \) suddenly decreases to a significantly lower value and eventually stabilizes.
This drop occurs due to the existence of bistable solutions in \( \overline{n}(N) \), analogous to the behavior observed when \( p \) varies. For larger \( p \) values (e.g., \( p = 0.6 \)), \( \overline{n} \) exhibits an sudden drop at smaller total qubit numbers \( N \) compared to cases with smaller \( p \). Thus, even if \( p \) remains constant, \( \overline{n} \) is influenced by the total number of qubits \( N \) in the system. Adjusting the total number of qubits \( N \) can enhance the mean excited-state population of the system.

One might inquire about the influence of qubit-qubit correlations on the aging transition.
To illustrate this effect, we derive a set of complex nonlinear equations (Eq.~(\ref{reply_eq30})) that are challenging to solve analytically.
However, numerical simulations provide valuable insight, with detailed calculations outlined in Appendix \ref{appendixE}.
From FIG.~\ref{reply_6}, it is evident that correlations lead to multiple aging transitions as \( p \) increases.
These correlations violate the conditions assumed in the mean-field approximation, resulting in a more intricate behavior of the aging transition.

\begin{figure}[t]
	\centering
	\includegraphics[width=0.43\textwidth]{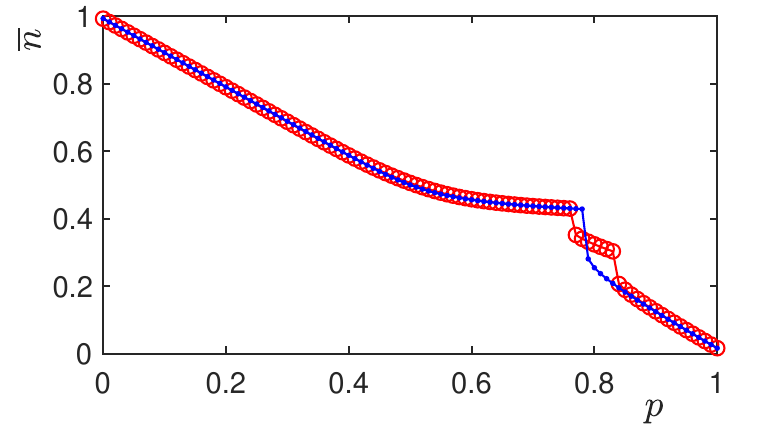}
	\caption{Excited-state population versus the ratio of inactive qubits. The circle line is for the results with second-order correlations. While the dotted line is for that without correlations. The parameters chosen are $N$=100, $\Delta$=3$\kappa$, $g$=0.04$\kappa$, $V=0$, and $\Omega$=3.2$\kappa$. The initial conditions are $\langle Q\rangle_0=\langle A\rangle_0=\langle B\rangle_0=\langle C\rangle_0=\langle D\rangle_0=\langle E\rangle_0$=0.}
	\label{reply_6}
\end{figure}

Before closing this section, it is worth noting that the model described by Eq.~(\ref{Hamiltonian}) and Eq.~(\ref{masterequation}) can be implemented using trapped ions \cite{Lee2013,Nadolny2023,Lorch2017,Hush2015}.
In this setup, each ion can function as a two-level system (qubit).
The linear gain and loss terms for active and inactive qubits can be engineered using blue and red sideband transitions, which incoherently excite and de-excite the qubits.
Previous studies have shown that both superconducting flux qubit systems \cite{John2011} and trapped-ion setups \cite{Zhang2017} can provide feasible experimental platforms to realize the model discussed in this manuscript, leveraging advancements in current quantum technology.

\section{CONCLUSIONS}\label{sec3}

In this manuscript, we investigated the aging transition in a network of coupled qubits classified into active and inactive types based on their dissipation characteristics.
Our findings contribute to understanding the aging of coupled qubits in several key ways. Firstly, we observed a distinct aging transition characterized by an abrupt decrease in the mean excited-state population as the ratio of inactive qubits to total qubits, denoted as \( p \), exceeded the critical threshold \( p_c \). Secondly, we demonstrated that the aging transition point \( p_c \) can be shifted towards larger values under weaker coherent and dissipative couplings, as well as by increasing the laser driving strength \( \Omega \). This indicates enhanced robustness of the network against aging under weaker couplings and stronger driving conditions.
Thirdly, we highlighted the sensitivity of \( p_c \) to initial conditions under fixed system parameters, where \( p_{\text{cmin}} \) and \( p_{\text{cmax}} \), representing the minimum and maximum aging transition points, respectively, were determined using linear stability analysis. Additionally, we found that the region where aging transitions occur depends on the coupling strength \( V \) and laser driving strength \( \Omega \): initially expanding with increasing \( \Omega \) and \( V \), then contracting. Finally, we showed that for a fixed \( p \), varying the total number of qubits \( N \) can optimize the mean excited-state population \( \overline{n} \), suggesting that an appropriate \( N \) can maximize \( \overline{n} \) under given system parameters.

\section*{ACKNOWLEDGMENTS}\label{sec5}
This work is supported by National Natural Science Foundation of China (NSFC) under Grant No. 12175033. We thank S. L. Wu and Jianning Li for the helpful discussions and constructive comments on our manuscript.

\appendix

\section{DERIVATION OF THE EQUATIONS OF COLLECTIVE MOTION}\label{appendixA}
To obtain the equations of collective motion for the expectation value $\langle Q\rangle$=Tr$(Q\rho)$, we multiply Eq.~(\ref{masterequation}) by $Q$ and obtain,
\begin{eqnarray}
\nonumber
\dot{\langle Q\rangle}&=&\text{Tr}(Q\dot{\rho})
=\frac{1}{N}\text{Tr}(\sum_j\sigma_{+}^j\sigma_{-}^j\dot{\rho}).\\\nonumber
\end{eqnarray}
Plugging the master equation into the last equation, we find
\begin{eqnarray}
\dot{\langle Q\rangle}&=&\frac{-i\Omega}{2N}\sum_{j=1}^{N} (\langle\sigma_{+}^{j}\rangle-\langle\sigma_{-}^{j}\rangle)
-\frac{4V(N-1)}{N^2}\sum_{j=1}^{N}\langle\sigma_{+}^{j}\sigma_{-}^{j}\rangle\nonumber\\&\ &+2\frac{\kappa}{N}\sum_{j=1}^{N_a}(1-\langle\sigma_{+}^{j}
\sigma_{-}^{j}\rangle)-2\frac{\kappa}{N}\sum_{j=N_a+1}^{N}\langle\sigma_{+}^{j}\sigma_{-}^{j}\rangle\nonumber\\
&\ &+\frac{2V}{N^2}{\sum_{j,k}}^{'}(\langle\sigma_{+}^{j} \sigma_{-}^{k}\rangle+\langle\sigma_{+}^{k}\sigma_{-}^{j}\rangle).
\label{Q}
\end{eqnarray}
From Eq.~(\ref{Q}), a new collective operator $\frac{1}{N}\sum_{j}\sigma_{+}^{j}$  appears. In order to get a closed set of equations, we define $A=\frac{1}{N}\sum_{j}\sigma_{+}^{j}$. Similarly, we can derive the equation for $\dot{\langle A\rangle}$,
\begin{eqnarray}
\label{A}
\dot{\langle A\rangle} &=&\frac{i[\Delta-4g(N-1)]-\kappa}{N}\sum_{j}\langle \sigma_{+}^{j}\rangle\nonumber\\
&\ &-\frac{i\Omega}{2}(\frac{2}{N}\sum_{j=1}^{N}\langle \sigma_{+}^{j}\sigma_{-}^{j}\rangle-1)\\
&\ &+\frac{4ig}{N}{\sum_{j,k}}^{'}2\langle\sigma_{+}^{j} \sigma_{+}^{k}\sigma_{-}^{k}\rangle
-\frac{4V}{N^2}{\sum_{j,k}}^{'}\langle\sigma_{+}^{j}\sigma_{-}^{j}\sigma_{+}^{k}\rangle.\nonumber
\end{eqnarray}
Simplifying Eq.~(\ref{Q}) and Eq.~(\ref{A}), we attain Eq.~(\ref{collective operator1}) and Eq.~(\ref{collective operator2}) in the maintext.

\section{COMPARISON OF THE RESULTS GIVEN BY THE EQUATIONS OF COLLECTIVE MOTION, MEAN-FIELD THEORY, AND NUMERICAL SIMULATIONS}\label{appendixB}
\begin{figure}[H]
	\centering
	\includegraphics[width=0.45\textwidth]{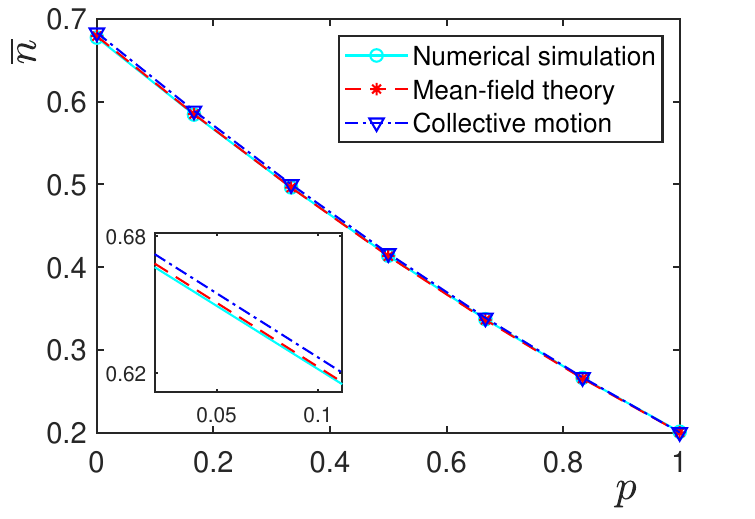}
	\caption {Comparison of results from the three approaches: equations of collective motion, the mean-field theory, and the numerical simulation. The line marked with circles represents the results by numerical simulation  without any approximations. The line marked with triangles represents the results given by the equations of collective motion and the line marked with stars represents the result obtained approximately by the mean-field theory. $N$=6 are chosen for this plot. Other parameters chosen are the same as in FIG.~\ref{aqbpic1}.}
	\label{aqbpic9}
\end{figure}

In this appendix, we demonstrate the reasonableness of the approximations made in the maintext by comparing them with the exact results obtained through numerical simulations based on Eq.~(\ref{masterequation}), as shown in FIG.~\ref{aqbpic9}. We choose a small  number of qubits, $N$=6, to illustrate the difference among the three approaches. From the figure, it can be observed that the results based on the equations of collective motion converge closely to the exact value when $p$ is small, and are almost in agreement when $p$ is large. For the mean-field approach, it is almost identical to the exact value from the beginning. This verifies the validity of the approaches presented in this manuscript. Due to the  substantial
computational effort associated with simulating large numbers of qubits, the comparison here is limited to the case of a small number of qubits.
\section{THE DIFFERENCE BETWEEN THE RESULTS GIVEN BY EQUATIONS OF COLLECTIVE MOTION AND MEAN-FIELD THEORY}\label{appendixC}
As shown in FIG.~\ref{aqbpic234}(a), a tiny difference exists in the critical point $p_c$ predicted by the  mean-field theory and the equations of collective motion. This difference can be understood as follows. By carefully sorting out Eq.~(\ref{wj}) with Eq.~(\ref{zhankaisuoxie1}), we observe three additional terms \( M_1 \), \( M_2 \), and \( M_3 \)  given by the equations of collective motion
\begin{eqnarray}
\begin{aligned}
M_1 &= i\frac{8g}{N}\sum_j{\langle\sigma_{+}^{j}\rangle}\langle\sigma_{+}^{j}\sigma_{-}^{j}\rangle,\\
M_2 &= \frac{4V}{N^2}\sum_j{\langle\sigma_{+}^{j}\rangle}\langle\sigma_{-}^{j}\rangle,\\
M_3 &= -\frac{4V}{N^2}\sum_j{\langle\sigma_{+}^{j}\rangle}\langle\sigma_{+}^{j}\sigma_{-}^{j}\rangle.
\end{aligned}
\end{eqnarray}
Particular attention should be paid to the case when \( V = 0 \). In this case, \( M_2 = M_3 = 0 \).
The difference between the two methods remains unchanged, indicating that the only term causing the difference is \( M_1 \).
We also verify that the presence of \( M_2 \) or \( M_3 \) does not affect the results even if \( V \neq 0 \).
Thus, the reason for the difference is the term \( M_1 \) in the equations of collective motion.
\section{THE STABILITY ANALYSIS FOR FIXED POINTS}\label{appendixD}
In this section, we provide a detailed analysis of the stability of the fixed points.
In FIG.~\ref{aqbpic1}, the fixed points of $\overline{n}$ is plotted as a function of $p$. From the figure, we observe that only one fixed point, $\overline{n}_1$, exists in region \uppercase\expandafter{\romannumeral1} and \uppercase\expandafter{\romannumeral3}, as there is only one real root to the cubic polynomial in Eq.~(\ref{Qeq}) in these two regions. In region \uppercase\expandafter{\romannumeral2}, there are three possible solutions   corresponding to the three real roots of the cubic polynomial. We find that, in the region \uppercase\expandafter{\romannumeral2}, $\overline{n}$ is allowed to transition from $\overline{n}_1$ to $\overline{n}_2$, as shown in FIG.~\ref{aqbpic234}(a) ($V$=0.2). An intriguing question arises: can we achieve a jump to $\overline{n}_3$ by changing the initial conditions \( {\langle Q\rangle}_0 \) and \( {\langle A\rangle}_0 \)? This question can be addressed using linear stability analysis \cite{Strogatz1994} for Eq.~(\ref{zhankaisuoxie1}). For the sake of simplicity, let $Z_j$ denote the $j$th fixed point of the parameter $\overline{n}(p)$ [see Eq.~(\ref{roots})], and $z_j (j=1,2,3)$ represent the corresponding fixed point of $\langle A\rangle$. Consider $\eta_1$ as a small perturbation away from $Z_j$, and $\eta_2, \eta_3$ as small perturbations away from $z_j$ and $z_{j}^{*}$, respectively. With these notations, the linearized equations are given by,
\begin{widetext}
\begin{eqnarray}
\dot{\eta_1}&=&-i\frac{\Omega}{2}(\eta_2-\eta_3)-
(2\kappa+\frac{4V(N-1)}{N})\eta_1+4V(z_{j}^{*}\eta_2+z_j\eta_3),\nonumber\\
\dot{\eta_2}&=&[i[\Delta-4g(N-1)]-\kappa]\eta_2-
i\Omega\eta_1+i8gN(z_j\eta_1+Z_j\eta_2)-4V(z_j\eta_1+Z_j\eta_2),\nonumber\\
\dot{\eta_3}&=&[-i[\Delta-4g(N-1)]-\kappa]\eta_3
+i\Omega\eta_1-i8gN(z_{j}^{*}\eta_1+Z_j\eta_2)-4V(z_{j}^{*}\eta_1+Z_j\eta_3).
\label{stableanalys1}
\end{eqnarray}
\end{widetext}

\begin{figure}[t]
	\centering
	\includegraphics[width=0.5\textwidth]{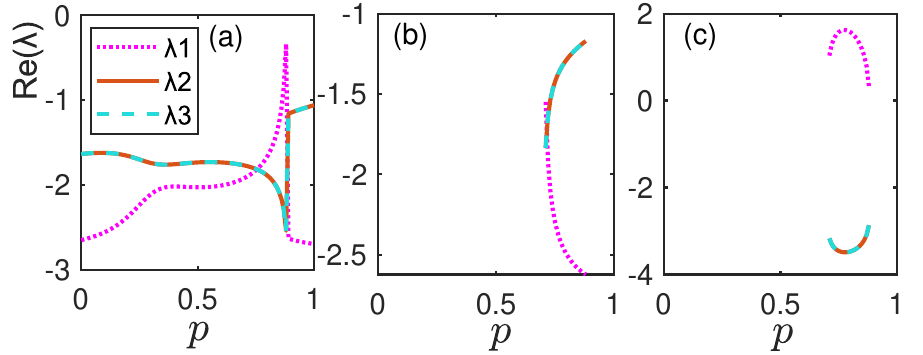}
	\caption{There are three fixed points between $p$=0.71 and 0.88. Figure (a), (b), and (c), respectively, corresponds to the real parts of three eigenvalues of matrix $M$ (defined in the Appendix) connected to the fixed points $\overline{n}_1$, $\overline{n}_2$, and $\overline{n}_3.$ The parameters chosen are the same as in FIG.~\ref{aqbpic1}.}
	\label{aqbpic5}
\end{figure}

This equation can be rewritten as,

\begin{equation}
\begin{pmatrix}
\dot{\eta_1} \\
\dot{\eta_2} \\
\dot{\eta_3}
\end{pmatrix}
=M
\begin{pmatrix}
\eta_1 \\
\eta_2 \\
\eta_3
\end{pmatrix}
\end{equation}
with $M$ is defined by,
\begin{widetext}
\begin{equation}
M=
\begin{pmatrix}
-(2G_j+\frac{4V(N-1)}{N}) & -i\frac{\Omega}{2}+4Vz_{j}^{*} & i\frac{\Omega}{2}+4Vz_{j} \\
-i\Omega+(i8gN-4V)z_j & i[\Delta-4g(N-1)]-\kappa+(i8gN-4V)Z_j & 0 \\
i\Omega-(i8gN+4V)z_{j}^{*} & 0 & -i[\Delta-4g(N-1)]-\kappa-(i8gN+4V)Z_j.
\end{pmatrix}
\end{equation}
\end{widetext}
By applying the ansatz $\eta(t)=e^{\lambda t}$ to Eq.~(\ref{stableanalys1}), the characteristic equation governing the stability of the fixed points is given by $|M-\lambda I|=0$. It is evident that $M$ has three eigenvalues $\lambda_1,\lambda_2,\lambda_3$, and the fixed point is stable if and only if the real parts of all  eigenvalues are negative. As seen in FIG.~\ref{aqbpic5}(a), the fixed point $\overline{n}_1$ remains stable as the real parts of its three eigenvalues are negative when $p$ ranges from 0 to 1. This ensures that $\overline{n}_1$ always exists regardless of how the fraction of inactive qubits changes. For $\overline{n}_2$, the results in FIG.~\ref{aqbpic5}(b) show that the real parts of all three eigenvalues are negative, indicating that $\overline{n}_2$ is stable in region \uppercase\expandafter{\romannumeral2}. This explains why it allows the transition from $\overline{n}_1$ to $\overline{n}_2$. However, for $\overline{n}_3$, it is clear from FIG.~\ref{aqbpic5}(c) that the real part of $\lambda_1$ is always positive in region \uppercase\expandafter{\romannumeral2}, violating the stability condition for fixed points. Therefore, $\overline{n}_3$ is not a stable fixed point.
\section{DERIVATION OF THE EQUATIONS OF COLLECTIVE MOTION IN THE PRESENCE OF SECOND-ORDER CORRELATION}\label{appendixE}
To take the second-order correlations into account, we define the operators,
$$B=\frac{1}{N}\sum_{j}\sum_{k\neq j}\sigma_{+}^{j}\sigma_{-}^{k},$$
$$C=\frac{1}{N}\sum_{j}\sum_{k\neq j}\sigma_{+}^{j}\sigma_{+}^{k}\sigma_{-}^{k},$$
$$D=\frac{1}{N}\sum_{j}\sum_{k\neq j}\sigma_{+}^{j}\sigma_{+}^{k},$$
$$E=\frac{1}{N}\sum_{j}\sum_{k\neq j}\sigma_{+}^{j}\sigma_{-}^{j}\sigma_{+}^{k}\sigma_{-}^{k}.$$
To simplify the problem, here we do not consider the existence of the dissipative coupling. Utilizing the relationship $\dot{\langle \mathcal{O}\rangle}$=Tr($\dot{\rho}\mathcal O$), the equations of motion for the collective operators are given by,
\begin{equation}
\begin{aligned}
\dot{\langle B\rangle}&=i\Omega(\langle C\rangle-{\langle C\rangle}^\ast)\nonumber
-\frac{i\Omega(N-1)}{2}(\langle A\rangle-{\langle A\rangle}^\ast)-2\kappa{\langle B\rangle},\\ \nonumber
\dot{\langle C\rangle}&=\frac{i\Omega (N-1)}{2}\langle Q\rangle+2\kappa N(1-p)\langle A\rangle+\frac{i\Omega}{2}\langle B\rangle\\&\ +(i\Delta+4ig-3\kappa)\langle C\rangle-\frac{i\Omega}{2}\langle D\rangle-i\Omega \langle E\rangle\\ \nonumber
&\ +\frac{4ig}{N}\sum_j\sum_k\sum_l\langle\sigma_{+}^{j}\sigma_{+}^{k}\sigma_{-}^{k}\sigma_{z}^{l}\rangle,\\ \nonumber
\dot{\langle D\rangle}&=i\Omega(N-1)\langle A\rangle-2i\Omega\langle C\rangle+(2i\Delta-2\kappa)\langle D\rangle\\ \nonumber
&\ +\frac{8ig}{N}\sum_j\sum_k\sum_l\langle\sigma_{+}^{j}\sigma_{+}^{k}\sigma_{z}^{l}\rangle,\\ \nonumber
\dot{\langle E\rangle}&=4\kappa N(1-p)\langle Q\rangle-i\Omega(\langle C\rangle-{\langle C\rangle}^\ast)-4\kappa\langle E\rangle.\nonumber
\label{reply_eq28}
\end{aligned}
\end{equation}
The calculation of these equations are similar to that of deriving the equations of the collective operator $\langle Q\rangle$ and $\langle A\rangle$, except that here we keep the two-qubit expectations  and approximate  the three-qubit expectations  with their cumulant approximations \cite{Kubo1962},
\begin{equation}
\begin{aligned}
\langle\sigma_{+}^{j} \sigma_{+}^{k}\sigma_{-}^{k}\sigma_{z}^{l}\rangle\rightarrow\langle\sigma_{+}^{j}\sigma_{+}^{k}\sigma_{-}^{k}\rangle\langle
\sigma_{z}^{l}\rangle+\langle\sigma_{+}^{j}\sigma_{z}^{l}\rangle\langle\sigma_{+}^{k}\sigma_{-}^{k}\rangle\nonumber\\
+\langle\sigma_{+}^{j}\rangle\langle\sigma_{+}^{k}\sigma_{-}^{k}\sigma_{z}^{l}\rangle-2\langle\sigma_{+}^{j}\rangle
\langle\sigma_{+}^{k}\sigma_{-}^{k}\rangle\langle\sigma_{z}^{l}\rangle,\nonumber\\
\langle\sigma_{+}^{j} \sigma_{+}^{k}\sigma_{z}^{l}\rangle\rightarrow\langle\sigma_{+}^{j}\sigma_{+}^{k}\rangle\langle\sigma_{z}^{l}\rangle+
\langle\sigma_{+}^{j}\sigma_{z}^{l}\rangle\langle\sigma_{+}^{k}\rangle\nonumber\\
+\langle\sigma_{+}^{j}\rangle\langle\sigma_{+}^{k}\sigma_{z}^{l}\rangle-2\langle\sigma_{+}^{j}\rangle
\langle\sigma_{+}^{k}\rangle\langle\sigma_{z}^{l}\rangle.\nonumber
\label{reply_eq29}
\end{aligned}
\end{equation}
With these considerations, we obtain  a closed set of nonlinear differential equations as follows,
\begin{equation}
\begin{aligned}
\dot{\langle Q\rangle}&=\Omega{\rm Im}\langle A\rangle-2\kappa\langle Q\rangle+2\kappa(1-p),\\
\dot{\langle A\rangle}&=\frac{i\Omega}{2}(1-2\langle Q\rangle)+[i\Delta-4ig(N-1)-\kappa]\langle A\rangle+8ig\langle C\rangle,\\
\dot{\langle B\rangle}&=i\Omega(\langle C\rangle-{\langle C\rangle}^\ast)-\frac{i\Omega(N-1)}{2}(\langle A\rangle-{\langle A\rangle}^\ast)-2\kappa{\langle B\rangle},\\
\dot{\langle C\rangle}&=\frac{i\Omega (N-1)}{2}\langle Q\rangle+2\kappa N(1-p)\langle A\rangle+\frac{i\Omega}{2}\langle B\rangle-\frac{i\Omega}{2}\langle D\rangle\\&\ -i\Omega \langle E\rangle+(i\Delta+4ig-3\kappa)\langle C\rangle+4ig[\langle C\rangle(2N\langle Q\rangle\\&\ -N+2)+2N\langle Q\rangle\langle C\rangle+2N\langle A\rangle\langle E\rangle-4N^2\langle A\rangle{\langle Q\rangle}^2],\\
\dot{\langle D\rangle}&=i\Omega(N-1)\langle A\rangle-2i\Omega\langle C\rangle+(2i\Delta-2\kappa)\langle D\rangle+8ig[\langle D\rangle\\&\ (2N\langle Q\rangle-N+2)+4N\langle A\rangle\langle C\rangle-4N^2{\langle A\rangle}^2\langle Q\rangle],\\
\dot{\langle E\rangle}&=4\kappa N(1-p)\langle Q\rangle-i\Omega(\langle C\rangle-{\langle C\rangle}^\ast)-4\kappa\langle E\rangle.
\label{reply_eq30}
\end{aligned}
\end{equation}
The initial conditions for  $\langle B\rangle, \langle C\rangle, \langle D\rangle$, \text{and}  $\langle E\rangle$ are denoted by $\langle B\rangle_0, \langle C\rangle_0, \langle D\rangle_0$, \text{and} $\langle E\rangle_0$. By numerically solving Eq.~(\ref{reply_eq30}), we can obtain the results with the second-order correlations.

\end{document}